# QUANTUM MEMORY PROTOCOL IN QED CAVITY BASED ON PHOTON ECHO

S.A.Moiseev

*Kazan Physical-Technical Institute of Russian Academy of Sciences, Kazan 420029 Russia*
E-mail: samoi@yandex.ru

A new protocol of the optical quantum memory based on the resonant interactions of the multi atomic system with a cavity light mode is proposed. The quantum memory is realized using a controllable inversion of the inhomogeneous broadening of the resonant atomic transition and impact interaction (on request) of additional short $2\pi$ - laser pulse resonant to an adjacent atomic transition. We demonstrate that the quantum memory protocol is effective for arbitrary storage time and can be used for new quantum manipulations with transient entangled states in the field-atoms evolution. The effect of the fast absorption and emission of the light field is predicted.

Photons are convenient carriers of quantum information [1] however realization of universal quantum memory (QM) for photons is still a difficult problem which attracts large attention due to its importance for quantum information science [1,2,3,4,5]. There are several proposals based on single atoms in an optical cavity [6] and on optical dense macroscopic media in free space [7,8,9]. First successful experiments with optically dense media have been made recently both with classical fields [10] and specific quantum states of light [11]. Optical QM includes delicate reversible unitary dynamics of the interacting light and medium. Controlling such dynamics in the macroscopic media opens a door for new investigations in quantum optics. Particularly the QM effect based on the electromagnetically induced transparency (EIT) was used recently for the controllable generation of spectrally narrow single photon fields [12]. Such QM technique was also proposed to stationary single- [13], two- and three color entangled light control [14] that looks prospective to quantum nondemolition measurements of single photon fields [15]. QM based on the photon echo technique [9] can be used to short light pulses [16] and effective multiply manipulations of single photon wave packets [17]. Storage and retrieval of the light states in the QM processes follow through the *transient entangled states* of the light and medium. Investigations and control of such quantum states are also important for understanding of the fundamental issues in the multi particle quantum dynamics.

In this paper a new QM protocol based on the interaction control in the multi atomic system (N>1, N is a number of atoms) and resonance mode of the quantum electrodynamics cavity is proposed. Using the proposed QM, the possibility of new quantum manipulations are shown for the reversible evolution of the field and multi atomic system with *arbitrary timescale* unlike QM technique based on photon echo [9] which was initially suggested for gases with Doppler broadened atomic transitions and than developed for solid state media [18,19,20,21]. In the proposed protocol, the quantum field of cavity mode is storied at the interaction with inhomogeneously broadened resonant transition of the multi atomic system. The two operations are used to control the reversibility in the quantum dynamics of field mode. The first operation includes a frequency inversion of the inhomogeneous broadening on the resonant atomic transition and the second procedure gives an additional $\pi$-phase kick to atomic states similar to the method of work [22] proposed for controlling the coherent dynamics of atom in the cavity QED. Analysis performed here have shown both the effective control of the reversible field-atom dynamics and demonstrated a new effect of *fast absorption and emission* at the field–atoms interaction. This method can be used for QM processes and analysis of unitary quantum dynamics in the more complicated multi-particle systems.

Let us consider the interaction between the field mode and multi-atomic system assuming that all the quantum evolution takes place within short enough temporal duration so $t < \Gamma^{-1}$ (where $\Gamma$ is a maximum decay constant in the field and atoms evolution). Model of the proposed QM protocol I discuss here is characterized by the N-atomic Jaynes-Cummings Hamiltonian added by the inhomogeneous broadening and interaction with an external short laser pulse:

$$\hat{H} = \hbar\omega_o \hat{a}^+\hat{a} + \sum_{j=1}^{N}\{\sum_{n=2}^{3} E_2^j \hat{P}_{nn}^j - \hbar(g_j \hat{a}\hat{P}_{21}^j + g_j^* \hat{a}^+ \hat{P}_{12}^j) - \tfrac{1}{2}\hbar\Omega(t)(e^{-i(\omega_{32}t+\varphi_j)}\hat{P}_{32}^j + e^{i(\omega_{32}t+\varphi_j)}\hat{P}_{23}^j)\},\quad(1)$$

where $\hat{P}_{mn}^j = |m_j\rangle\langle n_j|$ are the atomic operators, $\hat{a}$ and $\hat{a}^+$ are ladder operators of the resonant mode; $g_j$ is a coupling constant of j-th atom with a cavity mode, $\Omega(t)$ and $\omega_{32}$ are Rabi and carrier frequencies of the additional control laser pulse, $E_1^j = 0$, $E_{2,3}^j = E_{2,3} + \hbar\Delta_{2,3}^j$, where $E_2 = \hbar\omega_o$, the atomic detunings $\Delta_{2,3}^j$ are inhomogeneously broadened within a distribution $G_{2,3}(\Delta/\Delta_{n(2,3)})$ with spectral widths $\Delta_{n(2,3)}$.

Let it be that initially all the atoms are in the ground state and the probe field of the cavity mode is excited at time $t = 0$ so initial quantum state is $|\Psi(t=0)\rangle = |g\rangle \otimes |\psi_f\rangle$ (where $|g\rangle = \prod_{j=1}^{N}|1_j\rangle$, $|\psi_f\rangle = \sum_{n=0} C_n(0)|n\rangle$).

1) First of all I consider the case when the short control $2\pi$-pulse is applied with time delay $\tau$ after entrance of the probe field so $\Omega(t) \neq 0$ for $\tau < t < \tau + \delta t$. We evaluate Schrödinger equation

$$i\frac{d}{dt}|\varphi(t)\rangle = \hat{V}(t)|\varphi(t)\rangle \qquad (2)$$

corresponding to the representation $|\varphi(t)\rangle = \exp\{it[\omega_o \hat{a}^+\hat{a} + \sum_j(\omega_o \hat{P}_{22}^j + \hbar^{-1}E_3 \hat{P}_{33}^j)]\}|\Psi(t)\rangle$ so

$$\hat{V}(t<\tau) = \sum_{j=1}^{N}[\hat{V}_{(-)}^j + \Delta_3^j \hat{P}_{33}^j - \tfrac{1}{2}\Omega(t)(e^{-i\varphi_j}\hat{P}_{32}^j + e^{i\varphi_j}\hat{P}_{23}^j)], \qquad (3)$$

where $\hat{V}_{(\mp)}^j = \Delta_2^j \hat{P}_{22}^j \mp (g_j \hat{a}\hat{P}_{21}^j + g_j^* \hat{a}^+ \hat{P}_{12}^j)$ and also it was assumed $\Delta_2^j(t<\tau) = \Delta_2^j$. Using Eq.(3) and introducing the operators $\hat{T}_{(\mp)}[\tau] = \exp\{-i\tau \sum_{j=1}^{N}\hat{V}_{(\mp)}^j\}$ the solution of Eq. (2) for $t \leq \tau$ is obtained:

$$|\varphi(t)\rangle = \hat{T}_{(-)}[t]|\Psi(0)\rangle. \qquad (4)$$

Exact solution in Eq. (4) demonstrates the complete transference of energy and all the quantum information of the atomic system in the case if atomic frequencies $\omega_o + \Delta_2^j$ fill tightly the spectral region around the cavity mode frequency $\omega_o$ and if the average number of photons in the probe light is smaller than the total number of atoms $\langle \hat{n}\rangle < N$. In order to specify the important details in the quantum evolution of Eq. (2) I also evaluate a special case of single photon initial state $|\psi_f\rangle = |1\rangle$ where the wave function gets a well-known form

$$|\varphi_{(1)}(t)\rangle = c_{ph}(t)|\Psi(0)\rangle + \sum_{j=1}^{N} c_j(t)\hat{P}_{21}^j |g\rangle \otimes |0\rangle, \qquad (5)$$

with the following solution for a large enough time of interaction $t >> \Delta_{n(2)}$

$$c_{ph}(t) \cong e^{-\gamma t}, \qquad (6a)$$

$$c_j(t) \cong ig_j e^{-i\Delta_2^j t} \int_0^t dt' \exp[(i\Delta_2^j - \gamma)t'], \qquad (6b)$$

where $\gamma = \pi N |g_\Sigma|^2 G_2(0)$ and the following substitution $\sum_{j=1}^{N}|g_j|^2 \exp[i\Delta_2^j(t-t')] = N|g_\Sigma|^2 \tilde{G}_2[\Delta_{n(2)}(t-t')]$ (where $\tilde{G}_2[\Delta_{n(2)}(t-t')] = \int_{-\infty}^{\infty} d\Delta e^{i\Delta(t-t')}G_2[\Delta/\Delta_{n(2)}]$) has been used and it was supposed for simplicity that $\gamma << \Delta_{n(2)}$.

As seen from the Eqs. (6a) and (6b) the entangled state of the field and atoms in Eq. (5) transfers asymptotically to the pure atomic state ($c_{ph}(\gamma t \to \infty) \to 0$). Such *temporal arrow* is determined by the dephasing of atomic amplitudes in Eq.(6b) as well as in Eq.(4) by the evolution of general state $|\varphi(t)\rangle$. Universal QM needs a retrieval of the arbitrary initial field state $|\psi_f\rangle$ from any moment of time delay $\tau \sim \gamma^{-1}$ and $\tau >> \gamma^{-1}$. Two variants of QM protocol are demonstrated below to control the temporal arrow for both time scales though a strong similarity of field mode dynamics Eqs.(6a) and (6b) to the irreversible spontaneous decay of two level atom in free space.

To reverse the field-atoms quantum dynamics let us apply initially a $2\pi$ − laser pulse (where $\theta = \int_\tau^{\tau+\delta t} \Omega(t)dt = 2\pi$, $\delta t$ is a short pulse temporal) at $t = \tau$ that leads to the state:

$$|\varphi(\tau + \delta t)\rangle = \hat{U}(2\pi)|\varphi(\tau)\rangle = \hat{U}(2\pi)\hat{T}_{(-)}[\tau]\hat{U}^+(2\pi)\hat{U}(2\pi)|\Psi(0)\rangle = \hat{T}_{(+)}[\tau]|\Psi(0)\rangle, \qquad (7)$$

where $\hat{U}(2\pi) = \prod_{j=1}^{N}\hat{U}_j(2\pi)$, $\hat{U}_j(\theta = 2\pi) = \hat{P}_{11}^j - (\hat{P}_{22}^j + \hat{P}_{33}^j)$ for short resonant laser pulse where the influence of the atomic detunings $\Delta_{3,2}^j$ are negligible due to $|\Delta_3^j - \Delta_2^j|\delta t << 1$.

In the second step just after $2\pi$ − pulse we switch the atomic detunings so $\Delta_2^j(t > \tau + \delta t) = -\Delta_2^j(t<\tau) = -\Delta_2^j$. Such frequency inversion can be realized by several methods. The first method proposed [18] for solid state medium is based on the excitation by additional radio frequency $\pi$ − pulse tuned to a resonance with nuclei spins coupled with the atomic electrons through the strong hyperfine interaction.

Inversion of the nuclei spins signs changes the local hyperfine field on the electron spin leading to the reversion of the inhomogeneous broadening on the electron spin transition. Another method uses a switching of the electric field gradient in the resonant medium [19,20,21] and was experimented realized recently for rare-earth ions doping the crystals [21].

Let us introduce an operator $\hat{J}_{2\pi}(t)$ including the two operations: frequency inversion and laser $2\pi$ – pulse kick with negligible relative time delay. After $\hat{J}_{2\pi}(\tau)$ procedure we get a new Hamiltonian $\hat{V}(t > \tau + \delta t) = \sum_{j=1}^{N}[-\hat{V}_{(+)}^{j} + \hbar\Delta_3^j(t>\tau)\hat{P}_{33}^j]$ that determines the following evolution of the wave function for $t > \tau + \delta t$:

$$|\varphi(t)\rangle = \hat{T}_{(+)}[-(t-\tau-\delta t)]\hat{T}_{(+)}[\tau]|\Psi(0)\rangle. \qquad (8)$$

As seen from Eq. (8), $|\varphi(t = 2\tau + \delta t)\rangle = |\Psi(0)\rangle$ (where $\delta t \ll \tau$) which gives a complete reconstruction of the initial arbitrary state $|\Psi(0)\rangle$ in the echo signal without any detail assumptions about the specific spectral structure of the inhomogeneous broadening. However the perfect long-lived QM for photons needs the inhomogeneous broadening of the resonant atomic transition. It is significant note the unitary perfect reversibility of the field + atoms evolution in Eq. (8) which takes place for arbitrary time delays $\tau \sim \gamma^{-1}$ or $\tau \gg \gamma^{-1}$ between forward and backward quantum evolutions from arbitrary transient entangled state $|\varphi(\tau)\rangle$ of the field mode and atoms. This result principally extends the previous variant of the photon echo QM technique [9] which was proposed only for the case of a large enough time delay $\tau \gg \gamma^{-1}$ where the complete absorption of the initial probe field took place. It is obvious that the restored field should be touched at $t = 2\tau$ on demand since the field will be absorbed again at further evolution for $(t-2\tau) \gg \gamma^{-1}$, or we should apply later on a new $\hat{J}_{2\pi}(t \gg 2\tau)$ operation again. It is worth to note that after the complete absorption we can extend a lifetime of the stored quantum state transferring it to the long-lived atomic states similar to [9,18].

2) Using a single photon initial state let us analyze the second (simplified) version of the proposed QM including only spectral inversion of the inhomogeneous broadening (this procedure will be called by an operator $\hat{J}_0(t)$). In this case taking into account Eqs. (6a) and (6b) we get the following solution of wave function in Eq. (6) for $t > \tau + \delta t$ ($\delta t \to 0$):

$$c_{ph}(t) = 2e^{-\gamma t} - e^{-\gamma|t-2\tau|}, \qquad (9a)$$

$$c_j(t) = ig_j\{\exp[i\Delta_2^j(t-2\tau)]\int_0^\tau dt' \exp[(i\Delta_2^j - \gamma)t'] + \int_\tau^t dt' \exp[i\Delta_2^j(t-t')]c_{ph}(t')\}. \qquad (9b)$$

If $\gamma\tau \gg 1$ the solution in Eqs. (9a) and (9b) give a complete reconstruction of the light field for $t = 2\tau$ so $c_{ph}(2\tau)|_{\gamma\tau \gg 1} = -1$ and $c_j(2\tau)|_{\gamma\tau \gg 1} = 0$ (the field is absorbed again later on $c_{ph}(t \gg 2\tau) = 0$). It is easy to generalize this result transferring to the arbitrary initial field state $|\Psi(0)\rangle$. Here we have $|\varphi(\tau \gg \gamma^{-1})\rangle = \hat{T}_{(-)}[\tau]|\Psi(0)\rangle \cong |\Psi_{at}(\tau)\rangle \otimes |0\rangle$ assuming near complete absorption of the field. Taking into account $\Delta_2^j(t>\tau) = -\Delta_2^j$ for $t > \tau$ we rewrite the Hamiltonian of Eq. (3b) in the form $\hat{V}_{(+)}^j(t>\tau) = \Delta_2^j \hat{P}_{22}^j - (g_j \hat{b} \hat{P}_{21}^j + g_j^* \hat{b}^+ \hat{P}_{12}^j)$ introducing the field operators $\hat{b} = -\hat{a}$, $\hat{b}^+ = -\hat{a}^+$. Whereupon similar to Eq. (9) we get a perfect reconstruction of the field for $t = 2\tau$ to the following new state

$$|\varphi(2\tau)\rangle|_{\gamma\tau \gg 1} = \hat{T}_{(+)}[-\tau]|\Psi_{at}(\tau)\rangle \otimes |0\rangle = \sum_{n=0} C_n(0)(\frac{1}{\sqrt{n!}}\hat{a}^+ e^{i\pi})^n|0\rangle. \qquad (10)$$

The reconstructed wave function in Eq. (10) includes additional $\pi$ phase shift comparing to the initial state $|\Psi(0)\rangle$ that coincides with the result of the field reconstruction in free space obtained in [19] whereas the solution in the Eq. (8) includes $2\pi$ phase shift where an additional $\pi$ shift of the echo signal field is caught from new atomic state evolved after interaction with the $2\pi$-laser pulse.

Coming back to the solution of the Eq. (9) for small time delays $\gamma\tau \leq 1$ first of all we find an incomplete reconstruction of the field so the probability of the atomic excitation does not equal to zero for $t = 2\tau$: $P_{atoms}(2\tau) = \sum_{j=1}^{N}|c_j(2\tau)|^2 = 4e^{-2\gamma\tau}(1-e^{-2\gamma\tau})|_{\gamma\tau \leq 1} > 0$. This important distinction of these two variants of QM protocol is determined by different phase relations between two field terms in Eq. (9a): the first term is coupled with the single photon initial field whereas the second term is given by the echo signal field which has an additional $\pi$ phase shift. The sharpest difference of the simplified QM protocol takes place at time delay $\tau = \tau_D = \ln 2/(2\gamma)$ where the destructive effect between two interfering fields leads to the unpredictable fast absorption of the initial field. Putting $\tau = \tau_D$ in Eq. (9a) we find

$$c_{ph}(t; \tau = \tau_D) = 2e^{-\gamma t} - \tfrac{1}{2}e^{\gamma t} \quad \text{for } \tau_D < t < 2\tau_D; \quad 0 \quad \text{for } t > 2\tau_D, \tag{11}$$

whereas $P_{atoms}(\tau_D < t < 2\tau_D) = 3 - [4e^{-2\gamma t} + (1/4)e^{2\gamma t}]$ and $c_j(t) = \exp[i\Delta_2^j(t - 2\tau_D)]c_j(2\tau_D)$ and $P_{atoms}(t \geq 2\tau_D) = 1$ that means that the complete dephasing of atomic states reaches at $\tau = 2\tau_D$ when the field mode abruptly turns into the vacuum state. Thus inversion switching of the atomic frequencies at $\tau = \tau_D$ stimulates a fast absorption of photon which is completed at $t = 2\tau_D$. This interrupted evolution of light +atoms interaction demonstrates a dramatically opposite result in comparison with the QM dynamics given in Eq. (8). We add that the unusual dynamics in Eq. (11) can be restored in time by applying of two operations $\hat{J}_{2\pi}(t' > 2\tau_D)$ and $\hat{J}_0(t = 2t' - \tau_D)$ what leads to the reconstruction $|\Psi(2t')\rangle = |\Psi(0)\rangle$ of arbitrary initial state (fast absorption and emission of a single photon field in the QM protocol is shown in Fig.1.).

    In this paper, the new QM protocol is proposed for perfect retrieval of the arbitrary quantum state of the cavity field mode with short and long time delays. The protocol is based on controlling of interaction between single quantum mode and multi-atomic system using spectral inversion of inhomogeneous broadened resonant transition and applying on demand additional short $2\pi$ laser pulse at the adjacent atomic transition. Detail analysis of the quantum dynamics in the proposed protocol shows an unusual dynamical effect of the fast absorption (or emission) of the light field which is the result of the destructive quantum interference between the initial and irradiated fields evolved in the cavity. The proposed QM protocol opens new interesting opportunities for quantum manipulations with photons and entangled states of light + multi-atomic systems which have a special interests both for single photon fields and for relatively more intensive quantum light. Finally we note that the problems analyzed here have become close to the collapse and revival phenomenon [22] in the particular case of the negligible small inhomogeneous broadening. So the proposed QM protocol can be useful for the investigation of nonunitary irreversible decoherence processes in more complicated multi-particle media. Such phenomena are also interesting for analysis in terms of the general approach to quantum reversibility studied recently in the framework of generalized Loschmidt echo [23] In particularly using it looks interesting to analyze more general schemes of quantum reversibility with the QM protocol including new manipulations by the atomic detuning and coupling constant.

    This work was supported by the grant of Russian Foundation of Basic Research No. 06-02-16822-a.

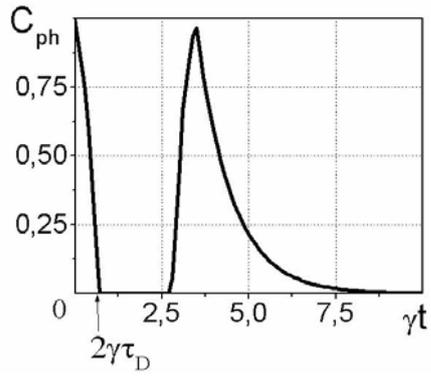

Figure 1. Quantum reversible dynamics of the field mode is shown for a single photon initial state at the following QM protocol: $\hat{J}_0(\tau_D)$, $\hat{J}_{2\pi}(5\tau_D)$, $\hat{J}_0(9\tau_D)$ ($\gamma\tau_D = \frac{1}{2}\ln 2 \cong 0.347$). As seen from the Fig. 1, there is fast absorption of field completed at $t = \tau_D$, which is restored then as a fast emission of the photon during $6\tau_D < t \leq 8\tau_D$.